\documentclass[a4paper]{article}

\usepackage{INTERSPEECH2021}
\usepackage{enumitem}
\usepackage{xspace}
\newcommand{\conformerS}{ConformerCTC 9M\xspace}
\newcommand{\conformerM}{ConformerCTC 28M\xspace}
\newcommand{\conformerL}{ConformerCTC 116M\xspace}
\newcommand{\conformerXL}{ConformerCTC 486M\xspace}
\newcommand{\conformerXXL}{ConformerCTC 631M\xspace}
\newcommand{\conformerA}{ConformerCTC 120M\xspace}
\newcommand{\conformerB}{ConformerCTC 582M\xspace}
\newcommand{\conformerC}{ConformerCTC 1017M\xspace}

\title{Pushing the Limits of Non-Autoregressive Speech Recognition}
\name{Edwin G. Ng$^1$\sthanks{\; Work done as a part of the Google AI Residency.}\;, Chung-Cheng Chiu$^1$, Yu Zhang$^1$\sthanks{\; Equal contribution.}\;, William Chan$^{2\dagger}$ }
\address{
  $^1$Google Research, Brain Team, United States\\
  $^2$Google Research, Brain Team, Canada}
\email{eg.ng@alum.utoronto.ca, \{chungchengc,ngyuzh,williamchan\}@google.com}

\begin{document}

\maketitle
\begin{abstract}
We combine recent advancements in end-to-end speech recognition to non-autoregressive automatic speech recognition. We push the limits of non-autoregressive state-of-the-art results for multiple datasets: LibriSpeech, Fisher+Switchboard and Wall Street Journal. Key to our recipe, we leverage CTC on giant Conformer neural network architectures with SpecAugment and wav2vec2 pre-training. We achieve 1.8\%/3.6\% WER on LibriSpeech test/test-other sets, 5.1\%/9.8\% WER on Switchboard, and 3.4\% on the Wall Street Journal, all without a language model.
\end{abstract}
\noindent\textbf{Index Terms}: speech recognition, non-autoregressive models

\section{Introduction}

End-to-end speech recognition models have been tremendously successful, and have achieved state-of-the-art results on various speech recognition tasks \cite{park2019specaugment}. However, much of recent advancements in WER have focused on autoregressive models such as LAS \cite{dong-icassp-2018} and RNN-T \cite{gulati2020conformer,zhang2020pushing}. There has been much recent work on non-autoregressive models as well \cite{chen-arxiv-2019,chan2020imputer,higuchi-interspeech-2020,higuchi-arxiv-2020}, however these models generally underperform the best autoregressive models. In this work, we are interested in pushing the limits of non-autoregressive speech recognition.

Deep neural networks have been observed to benefit from scaling model capacity \cite{krizhevsky2012imagenet,he2016deep,szegedy2015going}. Computer vision classification performance has benefited extensively from larger, deeper models \cite{krizhevsky2012imagenet,he2016deep,szegedy2015going}. Similarly, language models have benefited from very large Transformer networks \cite{vaswani2017attention,devlin2019bert,brown2020language}. Recent work in speech recognition have also observed this phenomena for end-to-end autoregressive speech recognition models \cite{gulati2020conformer,zhang2020pushing}.

Pre-training has significantly improved performance of neural networks for a multitude of tasks, for example natural language understanding \cite{wang2018glue,devlin2019bert}, object recognition \cite{chen2020big}, and speech recognition \cite{baevski2020wav2vec}. The idea of pre-training is to train a model with one task to learn useful parameters that can be transferred to other downstream tasks. For example, BERT \cite{devlin2019bert} has shown great success by using the Masked Language Model and Next Sentence Prediction pre-training tasks to achieve strong results on a variety of downstream NLP tasks such as the GLUE tasks \cite{wang2018glue}. Likewise, wav2vec pre-training \cite{baevski2020wav2vec} uses a similar pre-training task of masking encoded features and optimizing over a contrastive loss to obtain strong representations for the downstream task of automatic speech recognition.

A major challenge of training large models is due to their high capacities and ability to easily overfit the training data. Zhang et al. \cite{zhang2020pushing} show that scaling up to larger autoregressive speech recognition models without the use of wav2vec pre-training causes degradation in accuracy and that scaling up model size with wav2vec pre-training enables further performance gains on the LibriSpeech dataset \cite{panayotov2015librispeech}.

In this work, we combine recent techniques by training giant Conformer \cite{gulati2020conformer} models using wav2vec pre-training and a CTC decoder \cite{graves2014towards} to push the limits of accuracy for non-autoregressive ASR models. We achieve state-of-the-art non-autoregressive performance with WERs of 1.8\%/3.6\% on LibriSpeech \cite{panayotov2015librispeech} test-clean/test-other sets, 5.1\%/9.8\% on Switchboard/CallHome sets \cite{cieri2004fisher,godfrey1992switchboard}, and 3.4\% on the Wall Street Journal \cite{paul1992design} test92 set without utilizing an external language model.

The key contributions of this paper are summarized as follows:
\begin{enumerate}[noitemsep]
    \item We investigate the scaling of model sizes for non-autoregressive speech recognition.
    \item We investigate the importance of pre-training in very large models for non-autoregressive speech recognition.
    \item We show that by combining large parameter models with pre-training, we can push the limits of non-autoregressive speech recognition. We achieve SoTA on LibriSpeech, Fisher+Switchboard and Wall Street Journal compared to other non-autoregressive work.
\end{enumerate}

\begin{table*}[!ht]
  \caption{Configurations of our models. We utilize different configurations for our base models (without pre-training) vs. our pre-trained models (with wav2vec pre-training).}
  \label{tab:params}
  \footnotesize
  \centering
  \begin{tabular}{ l c c c c c c c }
    \toprule
    \textbf{Method} &
    \# Params (M) &
    Enc. Layers &
    Enc. Dim &
    Att. Head &
    Kernel Size &
    Batch Size \\
    \midrule
    \textbf{Base} \\
    \quad\conformerS & 8.9 & 16 & 144 & 4 & 32 & 2048 \\
    \quad\conformerM & 27.6 & 16 & 256 & 4 & 32 & 2048 \\
    \quad\conformerL & 115.7 & 17 & 512 & 8 & 32 & 2048 \\
    \quad\conformerXL & 485.6 & 18 & 1024 & 8 & 32 & 512 \\
    \quad\conformerXXL & 631.1 & 24 & 1024 & 8 & 32 & 512 \\
    \midrule
    \textbf{Pre-training} \\
    \quad\conformerA & 119.8 & 12 & 1024 & 8 & 5 & 256 \\
    \quad\conformerB & 581.7 & 24 & 2048 & 8 & 5 & 256 \\
    \quad\conformerC & 1016.5 & 42 & 2048 & 8 & 5 & 512 \\
    \bottomrule
  \end{tabular}
\end{table*}

\section{Related Work}
There has been much work done on scaling up models with convolution-based architectures \cite{li2019jasper,kriman2020quartznet,han2020contextnet,gulati2020conformer}. Jasper \cite{kriman2020quartznet} uses a series of blocks containing 1D convolutions, batch norm, ReLU, dropout and residual connections to enable deep variants of the network to be trained effectively (up to 54 convolution layers shown in the work) and QuartzNet \cite{kriman2020quartznet} further improves on the parameter efficiency by replacing the 1D convolutions with depthwise separable convolutions. ContextNet \cite{han2020contextnet} improves the encoding of global context information by adding squeeze-and-excitation modules \cite{hu2018squeeze} and also shows clear improvements to WER on LibriSpeech by scaling the model size from 11M to 113M parameters. Our work leverages the Conformer architecture \cite{gulati2020conformer}, which combines multi-headed self-attention \cite{vaswani2017attention} with convolutions to model local and global dependencies of the audio sequence in a parameter efficient way. The Conformer architecture has been observed to improve WER as we scale model capacity from 10M to 100M parameters \cite{gulati2020conformer}.

However, continuing to scale up model size does not enable continued improvements on performance. In fact, Zhang et. al \cite{zhang2020pushing} show that naively scaling up Conformer to model sizes of 600M and 1B actually degrades WER performance on LibriSpeech. However, by adding wav2vec pretraining, Zhang et. al \cite{zhang2020pushing} showed continued performance improvements for WERs on LibriSpeech as model sizes are scaled up and achieves state-of-the-art results for autoregressive speech recognition. In our work, we replace the Conformer RNN-T decoder with a CTC decoder and use the same combination of techniques as \cite{zhang2020pushing} without utilizing an external language model to achieve state-of-the-art results for non-autogressive speech recognition on LibriSpeech, Switchboard and Wall Street Journal.

\section{Method}
\subsection{Model Architecture: Conformer}
For the neural network architecture, we use a Conformer \cite{gulati2020conformer} encoder connected to a CTC decoder. The Conformer encoder is a convolution-augmented transformer which contains a stack of Conformer blocks, where each block has a series of feed forward layers, multi-headed self attention \cite{vaswani2017attention}, convolutions and layer norm. Like \cite{zhang2020pushing}, we remove the relative positional embedding \cite{dai2019transformer} from the self attention layer to speed up training for the models with wav2vec pre-training. Conformers of various sizes ranging from 9M to 1B parameters are trained from scratch without pre-training (base models) or with pre-training. The CTC decoder is a linear layer followed by a softmax function to project the encoder output to posterior probabilities. Details on specific architectures are shown in Table \ref{tab:params}.

\subsection{Wav2vec Pre-Training}
We pre-train the Conformer encoder using the variation of wav2vec in \cite{zhang2020pushing} with utterances from the "unlab-60k" subset of Libri-Light \cite{kahn2020libri}. Encoded features from the convolution subsampling layer are masked and fed into the Conformer encoder to produce context vectors. Target context vectors are obtained from running a linear layer on the encoded features and is compared with the context vectors from the encoder to optimize the contrastive loss.

\section{Experiments}
We use wav2vec (w2v) pre-trained Conformers with a CTC decoder to obtain SoTA non-autoregressive performance on Librispeech, Switchboard and Wall Street Journal test sets. We compare with other non-autoregressive models without a language model, since using a language model no longer makes the model end-to-end, and almost all practical language models are autoregressive. We however also include results with a language model from prior work for both non-autoregressive and autoregressive models for the readers convenience.

\subsection{Experiment Settings}
\label{pre_training_params}
\textbf{Data:} We follow the same procedure as \cite{zhang2020pushing} to pre-train our Conformer models with unlabeled audio from the Libri-Light \cite{kahn2020libri} dataset. We use 80-dimensional log-mel filter bank coefficients of the utterances
as single-channel input features for the networks. The pre-trained models are fine-tuned separately on two read speech datasets (i.e. 960h LibriSpeech and 80h Wall Street Journal) and one conversational speech dataset (2000h Fisher+Switchboard \cite{cieri2004fisher,godfrey1992switchboard}). \conformerS and \conformerB fine-tuned on Wall Street Journal use characters to tokenize the transcripts. All other models use a 1024-token subword vocabulary \cite{schuster2012japanese} constructed from the LibriSpeech 960h training set. More details comparing character versus subword tokenization are discussed in Section \ref{sec:tokenizer}.

\textbf{SpecAugment:} We use SpecAugment \cite{park2019specaugment} with two frequency masks using frequency mask parameter $F=27$ and ten adaptive time masks \cite{park2020specaugment} with a maximum time mask ratio of $p_s=0.05$ to augment the audio input for all models.

\textbf{Pre-training Parameters:} For masking, we follow the same parameters as \cite{baevski2020wav2vec} by sampling $p=0.065$ of all time steps to be starting indices and masking the subsequent $M=10$ time steps. We train with a global batch size of 2048 on 64/256/512 Google TPU V3 chips for 4-5 days for \conformerA, \conformerB and \conformerC models respectively. We use Adam optimization with a transformer learning rate schedule (section 5.3 of \cite{vaswani2017attention}) with 25k warm-up steps and peak learning rate 2e-3/2e-4 for \conformerA and \conformerB respectively. We apply gradient scaling to cap the norm of the gradient to 20. For \conformerC, we use Adafactor \cite{shazeer2018adafactor} with parameters $\beta_1=0.9$ and $\beta_2=0.98$ and a transformer learning rate schedule with peak learning rate 1e-3 and 25k warm-up steps.

\textbf{Fine-tuning Parameters:} We fine-tune the pre-trained checkpoints with global batch size 256/256/512 on 16/16/64 Google TPU V3 chips for 1-3 days for the \conformerA, \conformerB and \conformerC models respectively. The encoder and decoder are optimized with different learning rate schedules because the encoder was pre-trained separately. For \conformerA and \conformerB, we use Adam optimization with a transformer learning rate schedule with a peak learning rate of 3e-4 with 5k warm-up steps for the encoder and a peak learning rate of 8e-4 with 1.5k warm-up steps for the decoder. For \conformerC fine-tuned on LibriSpeech and Fisher+Switchboard, we use Adafactor with parameters $\beta_1=0.9$ and $\beta_2=0.98$ with a transformer learning rate schedule with a peak learning rate of 3e-4 with 5k warm-up steps for the encoder and a peak learning rate of 5e-4 with 1.5k warm-up steps for the decoder. For \conformerC fine-tuned on Wall Street Journal, we use the same parameters except we lower the encoder peak learning rate to 3e-5 and the decoder peak learning rate to 8e-5.

\textbf{Parameters for Base Models without Pre-training:} To better understand the effects of pre-training, we train from scratch five models of various sizes ranging from 9M to 631M parameters without pre-training. More details comparing pre-trained versus models trained from scratch can be found in Section \ref{sec:pre-training}. For all models trained from scratch, we train with global batch size 2048 on 64 Google TPU V3 chips for 1-4 days. We use Adam optimization with a transformer learning rate schedule with peak learning rate of 4e-3/3e-3/2e-3/1e-3/1e-3 with 10k warm-up steps for \conformerS, \conformerM, \conformerL, \conformerXL and \conformerXXL respectively.

\subsection{Experimental Results}
We present our results compared to previous non-autoregressive work and some non-autoregressive models which use an external LM in Table \ref{tab:librispeech_results}, Table \ref{tab:switchboard_results} and Table \ref{tab:wsj_results}. We find that character tokenization performs substantially better than subword tokenization for Wall Street Journal (more details in Section \ref{sec:tokenizer}). We also find that lowering the learning rate for \conformerC on Wall Street Journal, as described in Section \ref{pre_training_params}, is critical towards the optimization.

For LibriSpeech and Fisher+Switchboard (evaluated on Hub5'00), we find that our largest model \conformerC performs the best. For Wall Street Journal, the smallest of the three datasets, we find that our second largest model \conformerB performs the best. No language model is used in our models to allow the models to stay non-autoregressive. We hypothesize that because Wall Street Journal is a small dataset (80hrs compared to 960h LibriSpeech and 2000h Fisher+Switchboard), that larger models can easily overfit the training data, so a smaller learning rate is required for convergence which causes performance to degrade.

\begin{table}[ht]
  \caption{WERs (\%) for LibriSpeech. Our ConformerCTC models do not utilize an external language model, yet we outperform many prior work with strong language models.}
  \label{tab:librispeech_results}
  \footnotesize
  \centering
  \begin{tabular}{ @{} l c c c @{} }
    \toprule
    \textbf{Method} &
    \textbf{LM} &
    \textbf{test-clean} &
    \textbf{test-other} \\
    \midrule
    \textbf{Autoregressive} \\
    \quad LAS+SpecAugment \cite{park2019specaugment} & RNN & 2.5 & 5.8 \\
    \quad ContextNet \cite{han2020contextnet} & - & 2.1 & 4.6 \\
    \quad ContextNet \cite{han2020contextnet} & LSTM & 1.9 & 4.1 \\
    \quad Conformer \cite{gulati2020conformer} & - & 2.1 & 4.3 \\
    \quad Conformer \cite{gulati2020conformer} & LSTM & 1.9 & 3.9 \\
    \quad Conformer+w2v \cite{zhang2020pushing} & - & 1.6 & 3.3 \\
    \quad Conformer+w2v+NST \cite{zhang2020pushing} & Trans. & \textbf{1.4} & \textbf{2.6} \\
    \textbf{Semi-Autoregressive} \\
    \quad Imputer  \cite{chan2020imputer} & - & 4.0 & 11.1 \\
    \textbf{Non-Autoregressive} \\
    \quad DeepSpeech2 \cite{amodei2016deep} & 5-gram & 5.3 & 13.3 \\
    \quad wav2letter++ \cite{zeghidour2018fully} & ConvLM & 3.4 & 11.2 \\
    \quad Jasper \cite{li2019jasper} & - & 3.9 & 12.0 \\
    \quad Jasper \cite{li2019jasper} & T-XL & 2.8 & 7.8 \\
    \quad QuartzNet \cite{kriman2020quartznet} & - & 3.9 & 11.3 \\
    \quad QuartzNet \cite{kriman2020quartznet} & T-XL & 2.7 & 7.3 \\
    \quad CTC+w2v \cite{baevski2020wav2vec} & - & 2.2 & 4.5 \\
    \quad CTC+w2v \cite{baevski2020wav2vec} & Trans. & \textbf{1.8} & \textbf{3.3} \\
    \midrule
    \textbf{Base (Ours)} \\
    \quad \conformerL & - & 2.7 & 5.9 \\
    \textbf{Pre-training (Ours)} \\
    \quad \conformerC & - & \textbf{1.8} & \textbf{3.6} \\
    \bottomrule
  \end{tabular}
\end{table}

\begin{table}[ht]
  \caption{WERs (\%) for Switchboard.}
  \label{tab:switchboard_results}
  \footnotesize
  \centering
  \begin{tabular}{ l c c c}
    \toprule
    \textbf{Method} &
    \textbf{LM} &
    \textbf{SWB} &
    \textbf{CH} \\
    \midrule
    \textbf{Autoregressive} \\
    \quad Attention Seq2Seq \cite{weng2018improving} & - & 8.3 & 15.5 \\
    \quad RNN-T \cite{battenberg2017exploring} & 4-gram & 8.1 & 17.5 \\
    \quad Joint BPE \cite{wang2020investigation} & RNN & \textbf{4.9} & \textbf{9.5} \\
    \textbf{Non-Autoregressive} \\
    \quad CTC + Gram-CTC \cite{li2019jasper} & N-gram & 7.3 & 14.7 \\
    \quad Jasper \cite{li2019jasper} & T-XL & \textbf{7.8} & \textbf{16.2} \\
    \midrule
    \textbf{Base (Ours)} \\
    \quad \conformerL & - & 6.1 & 12.9 \\
    \textbf{Pre-training (Ours)} \\
    \quad \conformerC & - & \textbf{5.1} & \textbf{9.8} \\
    \bottomrule
  \end{tabular}
\end{table}

\begin{table}[ht]
  \caption{WERs (\%) for WSJ. Character tokenization significantly improves results compared to subword tokenization. See Section \ref{sec:tokenizer}.}
  \label{tab:wsj_results}
  \footnotesize
  \centering
  \begin{tabular}{l c c}
    \toprule
    \textbf{Method} &
    \textbf{LM} &
    \textbf{test92} \\
    \midrule
    \textbf{Autoregressive} \\
    \quad seq2seq + deep conv \cite{zhang2017very} & - & 10.5 \\
    \quad Joint BPE \cite{wang2020investigation} & RNN & 3.4 \\
    \quad Joint BPE + Stoc. Layer \cite{wang2020investigation} & RNN & \textbf{3.0} \\
    \textbf{Semi-Autoregressive} \\
    \quad Imputer  \cite{chan2020imputer} & - & 12.7 \\
    \quad Mask-CTC \cite{higuchi2020improved} & - & 9.1 \\
    \textbf{Non-Autoregressive} \\
    \quad wav2letter++ \cite{zeghidour2018fully} & 4-gram & 5.6 \\
    \quad wav2letter++ \cite{zeghidour2018fully} & ConvLM & \textbf{4.1} \\
    \quad Jasper \cite{li2019jasper} & - & 13.3 \\
    \quad Jasper \cite{li2019jasper} & T-XL & 6.9 \\
    \quad QuartzNet \cite{kriman2020quartznet} & 4-gram & 5.8 \\
    \quad QuartzNet \cite{kriman2020quartznet} & T-XL & 4.5 \\
    \midrule
    \textbf{Base (Ours)} \\
    \quad \conformerS (Char) & - & 11.8 \\
    \textbf{Pre-training (Ours)} \\
    \quad \conformerB (Char) & - & \textbf{3.4} \\
    \bottomrule
  \end{tabular}
\end{table}

\section{Discussion}
\subsection{Scaling Model Size and the Importance of Pre-training}
\label{sec:pre-training}
In this section, we will study the importance of model size and pre-training. We vary the model size and ablate pre-training. Table \ref{tab:librispeech}, Table \ref{tab:switchboard} and Table \ref{tab:wsj}, summarizes the results for LibriSpeech, Switchboard and WSJ.

On LibriSpeech we see that scaling up model size generally improves WER. However, we find that when we want to scale beyond 116M parameter models, we need pre-training to improve the model performance. Without pre-training, we found our models to overfit and generalize poorly. Pre-training allows us to scale to larger models and benefit from its capacity.

On Switchboard we see the exact same trend as LibriSpeech. Scaling up to our 116M parameter model improves WER, but beyond that pre-training is necessary for continual improvements. 

On Wall Street Journal we find that larger models worsen the WER. Pre-training significantly improves the WER and allows us to take advantage of the capacity of larger models, however scaling beyond the 582M parameter model also worsens the WER. We hypothesize that this different trend in results is due to the comparatively smaller training size of WSJ compared to the other two datasets. While Librispeech contains 960h of training audio and Fisher+Switchboard contains 2000h of training audio, WSJ only contains 80h of training audio. With the smaller training set, a high capacity model can easily overfit the data and damage generalizability in the test set. From these results, we see that pre-training allows one to exploit the capacity of larger models and that to continue improving performance of these larger models, the training data must also be sizable to prevent the issue of overfitting the training data. 

\begin{table}[ht]
  \caption{We compare different model sizes with and without pre-training (Base) on LibriSpeech using subword tokenization. No external language model is used. We observe that pre-training is needed to scale to very large models.}
  \label{tab:librispeech}
  \footnotesize
  \centering
  \begin{tabular}{ @{} l c c c c @{} }
    \toprule
    \textbf{Method} &
    \textbf{dev} &
    \textbf{dev-other} &
    \textbf{test} &
    \textbf{test-other} \\
    \midrule
    \textbf{Base} \\
    \quad \conformerS & 4.4 & 11.4 & 4.7 & 11.2 \\
    \quad \conformerM & 3.0 & 7.7 & 3.1 & 7.6 \\
    \quad \conformerL & \textbf{2.4} & \textbf{5.7} & \textbf{2.7} & \textbf{5.9} \\
    \quad \conformerXL & 2.7 & 6.2 & 2.8 & 6.3 \\
    \quad \conformerXXL & 2.8 & 6.9 & 3.0 & 6.9 \\
    \midrule
    \textbf{Pre-training} \\
    \quad \conformerA & 2.5 & 6.5 & 2.7 & 6.4 \\
    \quad \conformerB & 1.9 & 4.1 & 1.9 & 4.0 \\
    \quad \conformerC & \textbf{1.7} & \textbf{3.6} & \textbf{1.8} & \textbf{3.6} \\
    \bottomrule
  \end{tabular}
\end{table}

\begin{table}[ht]
  \caption{We compare different model sizes with and without pre-training on Switchboard using subword tokenization. No external language model is used. We observe that pre-training is needed to scale to very large models.}
  \label{tab:switchboard}
  \footnotesize
  \centering
  \begin{tabular}{l c c}
    \toprule
    \textbf{Method} &
    \textbf{SWB} &
    \textbf{CH} \\
    \midrule
    \textbf{Base} \\
    \quad \conformerS & 8.5 & 16.2 \\
    \quad \conformerM & 6.3 & 14.4 \\
    \quad \conformerL & \textbf{6.1} & \textbf{12.9} \\
    \quad \conformerXL & 7.9 & 16.4 \\
    \quad \conformerXXL & 6.5 & 15.0 \\
    \midrule
    \textbf{Pre-training} \\
    \quad \conformerA & 6.5 & 13.9 \\
    \quad \conformerB & 5.5 & 12.2 \\
    \quad \conformerC & \textbf{5.1} & \textbf{9.8} \\
    \bottomrule
  \end{tabular}
\end{table}

\begin{table}[ht]
  \caption{We compare different model sizes with and without pre-training (Base) on WSJ using subword tokenization. No external language model is used. Pre-training improves results, however degradation is observed in our largest model.}
  \label{tab:wsj}
  \footnotesize
  \centering
  \begin{tabular}{l c c}
    \toprule
    \textbf{Method} &
    \textbf{dev93} &
    \textbf{test92} \\
    \midrule
    \textbf{Base} \\
    \quad \conformerS & \textbf{22.9} & \textbf{19.4} \\
    \quad \conformerM & 34.9 & 30.5 \\
    \quad \conformerL & 40.9 & 36.2 \\
    \quad \conformerXL & 37.4 & 31.9 \\
    \quad \conformerXXL & 39.4 & 35.2 \\
    \midrule
    \textbf{Pre-training} \\
    \quad \conformerA & 8.5 & 5.7 \\
    \quad \conformerB & \textbf{6.5} & \textbf{4.4} \\
    \quad \conformerC & 7.1 & 5.8 \\
    \bottomrule
  \end{tabular}
\end{table}

\subsection{Tokenization}
\label{sec:tokenizer}
We compare the character tokenization versus the subword tokenization for Wall Street Journal. In Table \ref{tab:vocab}, we see the character tokenization improves substantially over the subword tokenization for Wall Street Journal. The Wall Street Journal corpus is substantially smaller than Switchboard and LibriSpeech, we hypothesize the subword tokenization is easier to overfit compared to the character tokenization. We also experimented with character tokenization for Switchboard and LibriSpeech, however in those cases we found the subword tokenization to outperform the character tokenization.

\begin{table}[ht]
  \caption{We compare two different tokenizations, subword vs. character, for WSJ. We find that character tokenization significantly outperforms subword tokenization on WSJ.}
  \label{tab:vocab}
  \footnotesize
  \centering
  \begin{tabular}{lc|cccc|cc|cc}
    \toprule
    \textbf{Method} &
    \textbf{Vocab} &
    \textbf{dev93} & 
    \textbf{test92} \\
    \toprule
    \textbf{Base} \\
    \quad \conformerS & Subword & 22.9 & 19.4 \\
    \quad \conformerS & Char &  14.0 & 11.8 \\
    \quad \conformerM & Subword & 34.9 & 30.5 \\
    \quad \conformerM & Char & 15.9 & 12.2 \\
    \quad \conformerL & Subword & 40.9 & 36.2 \\
    \quad \conformerL & Char & 21.0 & 16.7 \\
    \midrule
    \textbf{Pre-training} \\
    \quad \conformerA & Subword & 8.5 & 5.7 \\
    \quad \conformerA & Char & 6.4 & 4.6 \\
    \quad \conformerB & Subword & 6.5 & 4.4 \\
    \quad \conformerB & Char & \textbf{5.1} & \textbf{3.4} \\
    \quad \conformerC & Subword & 7.1 & 5.8 \\
    \quad \conformerC & Char & 4.8 & 4.0 \\
    \bottomrule
  \end{tabular}
\end{table}

\section{Conclusion}
We combine recent advancements in end-to-end speech recognition, and push the limits of non-autoregressive speech recognition. Key to our recipe, we leverage CTC on giant Conformer neural networks, SpecAugment and wav2vec2 pretraining. Our CTC models establish new state-of-the-art word error rates for non-autoregressive speech recognition on LibriSpeech, Switchboard and Wall Street Journal. Our work outperforms prior non-autoregressive work when compared without a external language model, and even outperforms many autoregressive prior work with strong language models, despite our work not using an language model. We achieve 1.8\%/3.6\% WER on LibriSpeech test-clean/test-other; on Switchboard, we achieve 5.1\%/9.8\% WER; and finally on Wall Street Journal we achieve 3.4\% WER.
\section{Acknowledgements}
We give thanks to Samy Bengio, Bo Li, and Anmol Gulati for reviewing our paper.

\bibliographystyle{IEEEtran}

\bibliography{main}

\end{document}